\documentclass[aps,prl,reprint,amsmath,amssymb,longbibliography]{revtex4-2}
\usepackage{graphicx,bm,booktabs}
\usepackage[colorlinks=true,allcolors=blue]{hyperref}
\newcommand{\F}{\mathcal F}
\newcommand{\B}{\mathcal B}
\newcommand{\fld}[1]{\mathsf{#1}}
\newcommand{\norm}[1]{\left\lVert#1\right\rVert}
\begin{document}
\title{Residual Carleman lift for nonlinear lattice Boltzmann simulations}
\author{Ilya Karlin}
\affiliation{Department of Mechanical and Process Engineering, ETH Zurich, CH-8092 Zurich, Switzerland}
\date{\today}
\begin{abstract}
Low-order Carleman embeddings replace nonlinear dynamics by a larger linear system, but truncation errors can accumulate even when the underlying nonlinear model is accurately approximated by a quadratic equation. We construct a residual Carleman lift (RC2) of lattice Boltzmann dynamics in which an independently evolved spectral background supplies the coefficients and forcing of a second-order residual lift. In a double-shear-layer test, conventional fixed-rest Carleman  reaches $87.3\%$ final error, whereas prepared RC2 achieves $0.136\%$ through vortex roll-up, without restarting the lift or using fine-reference feedback. The results demonstrate control of the finite-time accuracy of a fixed-order linear embedding through nonlinear preparation.
\end{abstract}
\maketitle

Nonlinear interactions govern the formation of structures and the transfer of energy across scales in fluids, plasmas, and many other physical systems. Their evolution nevertheless admits linear descriptions in enlarged spaces: observables, phase-space distributions, or products of the dynamical variables. Such embeddings preserve nonlinearity through the representation of the state, while transferring its evolution to a linear operator. Their renewed interest is partly motivated by quantum computation, whose linear state evolution makes linear embeddings natural candidates for simulating nonlinear classical physics~\cite{Joseph2020,Engel2021}.

The central difficulty is to turn an infinite linear representation into a useful finite computation. Carleman-based quantum algorithms have established conditional results for dissipative nonlinear differential equations~\cite{Liu2021}, with subsequent improvements through rescaling, higher-order numerical methods, and tighter error analysis~\cite{Costa2025}. However, finite-dimensional projection can introduce accumulating errors even when the underlying nonlinear dynamics remains well behaved~\cite{Lin2024}. The accuracy of the truncated embedding is therefore a physical and numerical question that precedes its quantum realization.

Carleman embedding evolves products of the original variables in a linear hierarchy. Lattice Boltzmann (LB) models provide a concrete setting: streaming is linear, while the local equilibrium carries the nonlinearity. Carleman--LB studies have demonstrated useful accuracy in selected regimes~\cite{Itani2022,Sanavio2025}. Together with general truncation estimates~\cite{Forets2017}, they motivate examining how the dynamics presented to a finite lift controls its accuracy horizon.

A quadratic nonlinear model generates cubic and quartic products in the discrete evolution of its second sector. Removing those products can destroy accuracy even when the quadratic model itself remains accurate. Increasing the embedding order enlarges the state. Here we investigate a different control: which part of the dynamics should be supplied by an independent nonlinear background before the residual is lifted?

We introduce a residual Carleman lift truncated at degree two (RC2). A time-dependent background supplies its coefficients and forcing, receiving no information from the residual or the fine nonlinear reference. The pair sector is initialized once and propagated independently; it is never reconstructed from the evolving first sector. This differs operationally from methods that reconstruct a Carleman embedding when switching local charts~\cite{Novikau2025}. Our objective is fidelity to nonlinear LB dynamics on the same lattice. Quantum realization and computational advantage are outside the present study.

\textit{Conventional Carleman lift.---}
Let $\fld f_n\in\mathcal V_N=\mathbb R^{9N^2}$ collect all postcollision populations on a periodic two-dimensional $N\times N$ lattice; $n$ denotes the time step. Its value at one node is $\fld f_n(\bm x)\in\mathbb R^9$. With weight vector $\fld w\in\mathbb R^9$, the uniform equilibrium at fixed density $\rho_0$ and zero velocity is
\begin{equation}
 \fld f^{(0)}=\rho_0\fld w\otimes\mathbf1_{N^2}.
 \label{eq:rest}
\end{equation}
It is independent of position and time. The full lattice update is $\fld f_{n+1}=\F(\fld f_n)$, so $\F(\fld f^{(0)})=\fld f^{(0)}$; its increment vanishes at equilibrium.

We use regularized D2Q9 dynamics~\cite{Latt2006} because its postcollision state has an exact six-field representation: density $\rho$, two momentum components $\bm j$, and three stress moments. It provides a compact nonlinear demonstrator. The residual-lift construction does not depend on this collision model, and accuracy is assessed within the stable operating regime of the underlying discrete dynamics. The population update and its six-field implementation are specified in the Supplemental Material~\cite{Supplement}.

\begin{figure}[t]
\includegraphics[width=\columnwidth]{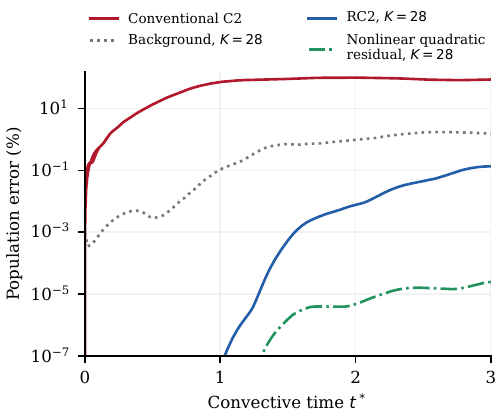}
\caption{Population errors relative to direct LB on $64^2$ sites in the double-shear layer simulation at $\mathrm{Re}=10^5$, with the common normalization in Eq.~\eqref{eq:error}. Curves show conventional fixed-rest C2, the prepared background at $K=28$, RC2 using that background, and the nonlinear quadratic residual solution at $K=28$. The green curve is obtained by evolving Eq.~\eqref{eq:nonlinearQ} nonlinearly, without Carleman lifting; its error reflects the accumulated omission of cubic and higher residual terms.}
\label{fig:errors}
\end{figure}

The conventional construction expands the local equilibrium about $(\rho_0,\bm j=0)$. Its quadratic term contains $\bm j\otimes\bm j/\rho_0$; density variations in the denominator first contribute at cubic order. Combining this expansion with collision and linear streaming gives, for $\fld x_n=\fld f_n-\fld f^{(0)}$,
\begin{equation}
 \fld x_{n+1}=A_*\fld x_n+\B_*[\fld x_n\otimes\fld x_n]
 +\mathcal R_{\geq3,*}(\fld x_n).
 \label{eq:conventionalmap}
\end{equation}
Here $A_*=D\F(\fld f^{(0)})$ is the Jacobian at the uniform equilibrium \eqref{eq:rest}, $\B_*$ contracts a global pair tensor and $\mathcal R_{\geq3,*}$ is the remainder of order three and higher. Streaming translations are included in these operators. Promoting products to independent sectors and truncating at degree two yields conventional second-order Carleman lift C2:
\begin{equation} \label{eq:conventionalC2}
\begin{aligned}
 \fld x_{n+1}&=A_*\fld x_n+\B_*[\fld Z_n],\\
 \fld Z_{n+1}&=(A_*\otimes A_*)\fld Z_n.
\end{aligned}
\end{equation}
The initial pair is $\fld Z_0=\fld x_0\otimes\fld x_0$. It includes populations at different nodes and subsequently evolves independently of $\fld x_n\otimes\fld x_n$. This conventional lift is constructed about the static background \eqref{eq:rest}. As the flow evolves away from that background, low-order truncation can accumulate substantial error, as in Fig.~\ref{fig:errors}.

\textit{Preparing the nonlinear background and lifting the residual.---} To address the limitations of the uniform background \eqref{eq:rest} for the Carleman lift, we aim at constructing a nonlinear background $\fld c_n$ and lift the \emph{residual} $\fld r_n=\fld f_n-\fld c_n$ instead of lifting $\fld x_n=\fld f_n-\fld f^{(0)}$. To that end, fix the flow setup, Reynolds number, a convective-time horizon $0\leq t^*\leq T^*$, and a lattice pair $(N_c,N_t)$, with $N_c<N_t$. The coarse lattice $N_c$ supplies the initial trajectory for preparation; the prepared background and residual lift both evolve on the target lattice $N_t$. In other words, our goal is to prepare a background that carries enough nonlinear evolution for a low-order Carleman \emph{residual} lift to remain accurate throughout the target interval $T^*$. The coarse trajectory on the $N_c$ grid alone need not provide such a background.

Let the coarse-grid populations $\fld g_m\in\mathcal V_{N_c}$ evolve by $\fld g_{m+1}=\F_c(\fld g_m)$ on $N_c^2$ coarse sites; $\F_t$ denotes the LB map on $N_t^2$ target sites. For common lattice velocity $U$, coarse and target times are $t^*=Um/N_c$ and $Un/N_t$, respectively. The required step counts are $M_c=T^*N_c/U$ and $M_t=T^*N_t/U$. Spatial prolongation $\mathcal P_{c\to t}$ and interpolation $\mathcal I_c$ of the coarse time series define the transferred coarse trajectory $\fld b_n$ on the target grid $N_t$:
\begin{equation}
 \fld b_n=\mathcal P_{c\to t}\mathcal I_c[\{\fld g_m\}](Un/N_t)
 \in\mathcal V_{N_t}.
 \label{eq:transfer}
\end{equation}
Thus $\fld b_n$ represents the coarse solution at target nodes and times; it is not evolved by $\F_t$.

To prepare the background $\fld c_n\in\mathcal V_{N_t}$, we evolve selected modes nonlinearly under the target map and prescribe their complement from $\fld b_n$:
\begin{equation} \label{eq:background}
\begin{aligned}
& \fld c_0=P_K\fld f_0+(I-P_K)\fld b_0,\\
& \fld c_{n+1}=P_K\F_t(\fld c_n)+(I-P_K)\fld b_{n+1}.
\end{aligned}
\end{equation}
Here the spectral projection $P_K$ retains integer Fourier wavevectors $|\bm k|\leq K$ in each target population field,
\begin{equation}
    \label{eq:projector}
    \widehat{P_K\fld h}(\bm k)=\left\{\begin{aligned}&\widehat{\fld h}(\bm k),\  k_x^2+k_y^2\le K^2\\
                                                &0,\ \text{otherwise}\end{aligned}\right.
\end{equation}
The projector retains the zero mode, so $P_K\fld f^{(0)}=\fld f^{(0)}$. It acts only in background preparation. The spectral cutoff $K$ will be specified below. 
With this, the nonlinear background $\fld c_n$ is constructed. While the background is intended to capture some essential nonlinearity of the flow, its remaining error can still exceed the desired accuracy, as in Fig.~\ref{fig:errors}.

We now introduce the residual $\fld r_n$ relative to the background by writing $\fld f_n=\fld c_n+\fld r_n$. Both the residual and its unrestricted pair sector evolve on the target lattice $N_t$. 
Importantly, by construction, the background $\fld c_n$ receives no feedback from the residual dynamics. 
The prescribed forcing is the one-step defect of the background:
\begin{equation}
 \fld d_n=\F_t(\fld c_n)-\fld c_{n+1}
 =(I-P_K)[\F_t(\fld c_n)-\fld b_{n+1}].
 \label{eq:defect}
\end{equation}
The defect \eqref{eq:defect} is an input to the lift, not an auxiliary diagnostic: even a zero residual evolves to $\fld d_n$ at the next step. 
Indeed, expanding the exact identity, $\fld r_{n+1}=\F_t(\fld c_n+\fld r_n)-\fld c_{n+1}$, gives a closed equation for the residual,
\begin{align}
 \fld r_{n+1}={}&\fld d_n+A_n\fld r_n+\B_n[\fld r_n\otimes\fld r_n]+\mathcal R_{\geq3,n}(\fld r_n),
 \label{eq:residual}
\end{align}
where $A_n=D\F_t(\fld c_n)$ and $\B_n[\fld h\otimes\fld k]=\tfrac12D^2\F_t(\fld c_n)[\fld h,\fld k]$. Similar to the conventional map \eqref{eq:conventionalmap}, the rational density dependence of the LB equilibrium gives an explicit higher-order remainder $\mathcal R_{\geq3,n}$~\cite{Supplement}. As in C2, RC2 evolves an independent symmetric pair tensor, now initialized from the residual: $\fld r_0=\fld f_0-\fld c_0$ and $\fld Z_0=\fld r_0\otimes\fld r_0$. Introduce linear combinations
\begin{equation}\label{eq:snqn}
\begin{aligned}
  &\fld s_n=A_n\fld r_n+\fld q_n,\\
  &\fld q_n=\B_n[\fld Z_n].
 \end{aligned}
\end{equation}
The RC2 system is
 \begin{align}
 \fld r_{n+1}&=\fld d_n+\fld s_n,\label{eq:c2r}\\
 \fld Z_{n+1}&=(A_n\otimes A_n)\fld Z_n+\fld d_n\otimes\fld s_n\notag\\
 &+\fld s_n\otimes\fld d_n+\fld d_n\otimes\fld d_n.
 \label{eq:c2z}
\end{align}
All terms of residual degree at most two are retained, treating the prescribed background forcing $\fld d_n$ \eqref{eq:defect} as degree zero, including its couplings to the lifted quadratic sector $\fld q_n$ \eqref{eq:snqn}. With the background prescribed, the RC2 equations \eqref{eq:c2r} and \eqref{eq:c2z} are affine linear in $(\fld r,\fld Z)$, or homogeneous linear after adjoining a constant. No products of the unknown first-sector state are used to close the pair sector. We propagate the RC2 lift classically through an exact factor representation, retaining all spatial pairs without rank truncation or reinitialization~\cite{Supplement}. For the uniform-background special case, replace the initialization in Eq.~\eqref{eq:background} by $\fld c_0=\fld f^{(0)}$ and set $\fld b_n=0$. Keeping the same projector and update gives 
$\fld c_{n+1}=P_K\F_t(\fld f^{(0)})=\fld f^{(0)}$ in \eqref{eq:background} and $\fld d_n=0$ in \eqref{eq:defect}, and RC2 \eqref{eq:c2r} and \eqref{eq:c2z} becomes the conventional C2 \eqref{eq:conventionalC2}.

For diagnostic purposes only, we separately evolve the nonlinear quadratic equation obtained by neglecting the remainder in the exact residual equation \eqref{eq:residual},
\begin{equation}
\tilde{ \fld {r}}_{n+1}=\fld d_n+A_n\tilde{\fld r}_n
 +\B_n[\tilde{\fld r}_n\otimes\tilde{\fld r}_n].
 \label{eq:nonlinearQ}
\end{equation}
With $\tilde{\fld r}_0=\fld r_0$, this auxiliary evolution uses the same background and actual residual products. It is not required to execute RC2 and supplies no feedback to it.
The comparison separates errors from the quadratic approximation of the residual dynamics from those introduced by truncating its Carleman lift; an exact identity for the latter's pair-consistency defect is derived in the Supplemental Material.

\textit{A coarse pilot and transfer of preparation.---} Finally, it remains to specify the spectral cutoff $K$.
The latter is evaluated using a pilot simulation construction. The pair of coarse-target lattices $(N_c,N_t)$ above is modeled by a surrogate pair $(N_c^\prime,N_t^\prime)$, with $N_t^\prime=N_c$, $N_c^\prime<N_t^\prime$. For the surrogate pair $(N_c^\prime,N_t^\prime)$, we first compute the nonlinear LB trajectories for the same flow and horizon, then screen predefined cutoffs to select $K^\prime$, which is transferred to the $(N_c,N_t)$ pair. In this work, we test a double shear layer in the periodic unit square at $\mathrm{Re}=10^5$, with lattice velocity $U=0.08$, layer width $0.04$, and transverse perturbation $0.05U\sin(2\pi x)$. The grid-dependent relaxation time $\tau_N=1/2+3UN/\mathrm{Re}$ preserves $U$ and Reynolds number. We prescribe $T^*=3$: 600, 1200, and 2400 steps on $N_c^\prime =16$, $N_t^\prime=N_c=32$ and $N_t=64$ lattices, respectively. These tests measure fidelity to the same-lattice nonlinear LB dynamics, rather than resolved continuum turbulence.
All population errors use
\begin{equation}
 \epsilon(n)=100\frac{\norm{\fld f_n^{\rm app}-\fld f_n^{\rm LB}}_w}{\norm{\fld f_0-\fld f^{(0)}}_w}.
 \label{eq:error}
\end{equation}
Here app specifies the method and $N$ is the lattice size used for that comparison.
The weighted norm is $\norm{\fld g}_w^2=\fld g^T W^{-1}\fld g$, where $W=\operatorname{diag}(\fld w)\otimes I_{N^2}$. It equals the Euclidean norm in the six orthonormal moment coordinates.

\begin{figure}[t]
\includegraphics[width=\columnwidth]{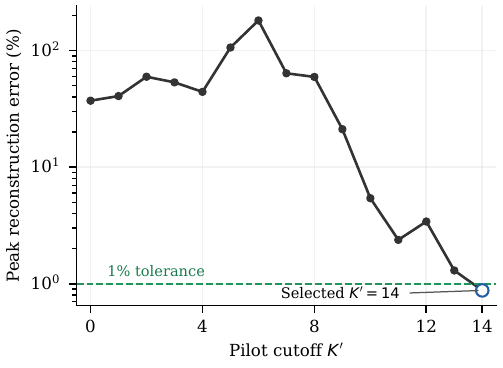}
\caption{Cutoff selection on the $(N_c^\prime,N_t^\prime)=(16,32)$ pilot at $\mathrm{Re}=10^5$. Each point is the peak population error through $t^*=3$ for a candidate prepared background with its linear residual correction. The dashed line is the prescribed $1\%$ tolerance; the open circle marks the smallest passing cutoff, $K^\prime=14$.}
\label{fig:preparation}
\end{figure}

The pilot pair $(N_c^\prime,N_t^\prime)=(16,32)$ uses shared nonlinear LB trajectories on $16^2$ and $32^2$ sites. For each integer $K^\prime=0,\ldots,14$, the candidate background and its \emph{linear} residual correction both evolve on the pilot target lattice, $32^2$, with $\fld r_{n+1}^{\rm lin}=\fld d_n+A_n\fld r_n^{\rm lin}$. The score is the maximum error of their reconstructed solution against the shared $32^2$ reference over the entire prescribed interval. The smallest cutoff meeting a $1\%$ tolerance is $K^\prime=14$ (Fig.~\ref{fig:preparation}). This screening step assesses preparation; it is not a bound on RC2 error.

We transfer this preparation to $(N_c,N_t)=(32,64)$ using $K/N_t=K^\prime/N_t^\prime$, preserving approximately the fraction of Fourier modes evolved nonlinearly. Its target accuracy is then tested over the prescribed interval. Both lattice sizes and the cutoff are doubled, giving $K=28$, with the fine-to-coarse ratio remaining two. The same flow and horizon $T^*$ are used; the choice is specific to both. The $N_t=64$ nonlinear reference is used only for validation, never for preparation or lifted evolution.

\begin{figure*}[t]
\includegraphics[width=0.82\textwidth]{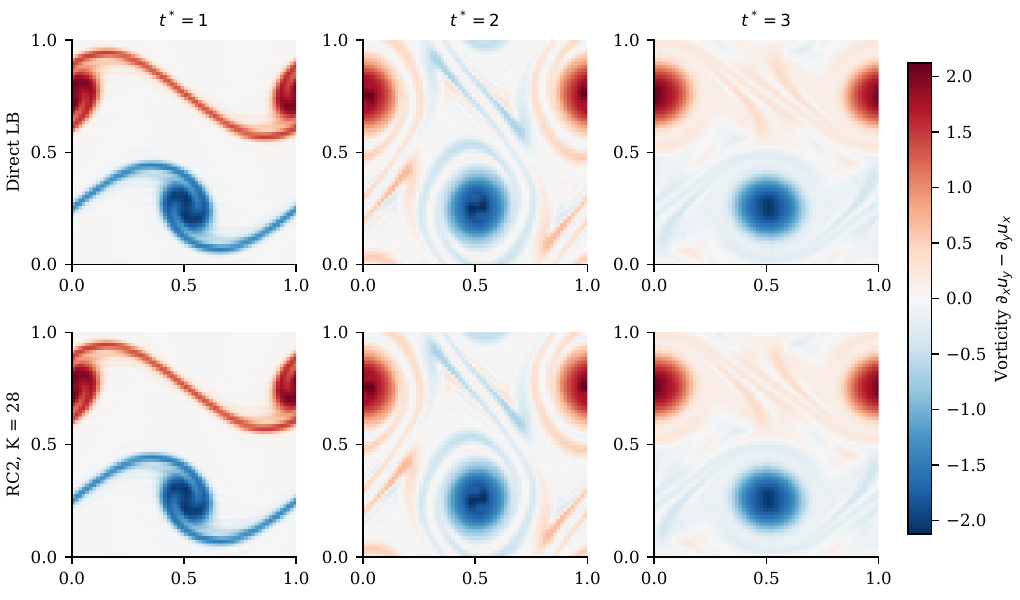}
\caption{Double-shear-layer roll-up at $\mathrm{Re}=10^5$ on a $64^2$ lattice. Direct nonlinear LB (top) and RC2 with $K=28$ preparation (bottom), at three convective times. All panels use the same reference-based vorticity scale and display unsmoothed lattice fields. The final relative vorticity error is $0.9045\%$. The RC2 lift is propagated for 2400 steps without reinitialization.}
\label{fig:rollup}
\end{figure*}

\textit{Accuracy through nonlinear roll-up.---}
Figure~\ref{fig:errors} compares the conventional C2 lift \eqref{eq:conventionalC2}, the prepared background \eqref{eq:background}, the residual RC2 lift \eqref{eq:c2r} and \eqref{eq:c2z}, and the nonlinear quadratic residual solution \eqref{eq:nonlinearQ}. For this $N_t=64$, ${\rm Re}=10^5$ test, the fixed-rest linearization (C1, not shown) first exceeds the $1\%$ population error after one lattice step, $t^*=0.00125$; its error subsequently oscillates. Conventional C2 first crosses $1\%$ at $t^*=0.15375$, after 123 lattice steps, and ends at $87.34\%$. At $K=28$, the prepared background has $1.539\%$ final error, RC2 has $0.1358\%$, and the nonlinear quadratic residual solution has $2.48\times10^{-5}\%$. The green curve thus measures the error accumulated by omitting cubic and higher terms of the exact residual equation, evolved nonlinearly; the blue curve is obtained by Carleman-lifting that quadratic equation and truncating its hierarchy at degree two. It is noted that, while the nonlinear background by itself is more accurate than the standard C2, the RC2 is a genuine improvement of the background. All four curves use the same LB reference and normalization.

The final RC2 velocity and vorticity errors, normalized by the corresponding current reference norms, are $0.1309\%$ and $0.9045\%$. Figure~\ref{fig:rollup} shows the formation and deformation of the vortices over the full interval. Additional cutoff sweeps and the diagnostic return of omitted cubic sources to the population sector are reported in the supplement. For the selected cutoff $K=28$, the prepared background retains $59.9\%$ of the target Fourier modes. RC2 reduces its final population error from $1.54\%$ to $0.136\%$, demonstrating that the remaining deviation can be corrected by a second-order linear lift over the prescribed interval.

Residual Carleman lifting is a hybrid construction: a small nonlinear evolution supplies the coefficients and forcing of a larger linear lifted evolution through one-way coupling. The coarse pilot serves as a numerical sandbox, rehearsing the flow over the prescribed interval of convective time and selecting the nonlinear spectral fraction to retain. That preparation is then transferred to the larger lattice. For the tested shear layer, it markedly reduces error accumulation through nonlinear roll-up at unchanged lifting order and without restarting the lift. The construction applies beyond LB; its usefulness depends on preparing a residual that a finite lift can evolve accurately over the required time interval.

\paragraph{Acknowledgement of AI assistance.}
During the development and preparation of this manuscript, the author used
ChatGPT (OpenAI, GPT-5.6 Sol) as an interactive research and writing assistant.
All results and final text were reviewed and accepted by the
author, who assumes full responsibility for the content of the manuscript.

\end{document}